# Structural and Vibrational Properties of the TiOPc monolayer on Ag(111)

Laura Fernandez[1], Sebastian Thussing[1], Alexander Mänz[1], Gregor Witte[1], Anton X. Brion-Rios[2,3], Pepa Cabrera-Sanfelix[2,4], Daniel Sanchez-Portal[2,5], and Peter Jakob[1]*

[1] Fachbereich Physik und Wissenschaftliches Zentrum für Materialwissenschaften der Philipps - Universität Marburg, Renthof 5, 35032 Marburg, Germany
[2] Donostia International Physics Center (DIPC), Paseo Manuel de Lardizabal 4, 20018 San Sebastián, Spain
[3] Departamento de Física de Materiales, UPV/EHU, Paseo Manuel de Lardizabal 3, 20018 San Sebastián, Spain
[4] IKERBASQUE, Basque Foundation for Science, Bilbao E-48011, Spain
[5] Centro de Física de Materiales CSIC-UPV/EHU, Paseo Manuel de Lardizabal 5, 20018 San Sebastián, Spain

## Abstract

The evolution of titanyl-phthalocyanine (TiOPc) thin films on Ag(111) has been investigated using IRAS, SPA-LEED and STM. In the (sub)monolayer regime various phases are observed that can be assigned to a 2D gas, a commensurate and a point-on-line phase. In all three phases the non-planar TiOPc molecule is adsorbed on Ag(111) in an oxygen-up configuration with the molecular $\pi$-conjugated backbone oriented parallel to the surface. The commensurate phase reveals a high packing density, containing two molecules at inequivalent adsorption sites within the unit cell. Both molecules assume different azimuthal orientations which is ascribed to preferred sites and azimuthal orientations with respect to the Ag(111) substrate and, to a lesser extent, to a minimization of repulsive Pauli interactions between adjacent molecules at short distances. At full saturation of the monolayer the latter interaction becomes dominant and the commensurate long range order is lost. DFT calculations have been used to study different adsorption geometries of TiOPc on Ag(111). The most stable configurations among those with pointing up oxygen atoms (bridge$_+$, bridge$_x$, top$_x$) seem to correspond to those identified experimentally. The calculated dependence of the electronic structure and molecular dipole on the adsorption site and configuration is found to be rather small.

## 1. INTRODUCTION

Titanylphthalocyanine (TiOPc) is a nonplanar π-conjugated molecule with pyramidal structure that exhibits exceptional absorption properties in the visible and near-infrared region.[1,2] It is one of the most sensitive organic photoconductors[3] and therefore its optical and electronical properties have been studied extensively.[4-9] For TiOPc bulk or thin film samples various polymorphic structures with particular stacking modes exist which can be distinguished by their different conductivity, mobility and optical properties.[2,7] In order to successfully prepare crystalline molecular thin films on surfaces with a uniform and well defined internal structure, it is essential to attain an in-depth understanding and good control of the relevant growth parameters. This includes in particular a thorough characterization of the initial growth conditions, i.e. the adsorption geometry of produced nucleation seeds and relevant adsorption geometries, as well as a clarification of the role of the substrate in the stacking of the molecules.

In the particular case of TiOPc the growth and electronic properties of TiOPc monolayers and ultrathin films have been investigated for a variety of substrates such as Au, Ag or graphite.[9-12] In most of these studies it has been concluded from scanning tunneling microscopy (STM) or Penning ionization electron spectroscopy (PIES) that the deposited TiOPc molecules adopt a planar adsorption geometry and, in the monolayer regime, maintain their 4 fold symmetry. In the literature only one noteworthy exception has been reported which is the growth of TiOPc on Ag(111). For this system the TiOPc was reported to adopt a strongly inclined molecular orientation.[11] Thereby, the significantly reduced substrate surface contact associated with such an arrangement suggests a negligible molecular-substrate interaction as compared to the intermolecular interaction, which the authors of that study considered as the most important parameter influencing TiOPc thin film growth.

This finding would imply a completely new scenario regarding the interaction and growth behavior of phthalocyanine molecules on coinage metal surfaces. In order to clarify this controversial observation we have performed an extensive study by combining spot profile analysis low energy electron diffraction (SPA-LEED), STM and infrared absorption spectroscopy (IRAS). Thereby we provide a revised model of TiOPc on Ag(111) regarding the adsorption geometry, molecular orientation and long range order of the TiOPc/Ag(111) monolayer system.

Because of the polarization of vibrational modes with respect to the molecular frame, IRAS can be used to derive precise information on molecular orientations of highly symmetric molecular species. Of particular interest in this regard are the intense TiOPc out-of-plane modes at 720 – 770 cm$^{-1}$, the Ti=O stretch at 960 cm$^{-1}$, as well as the 'in-plane' modes at >1000 cm$^{-1}$. The latter ones are very intense in the IR spectra of the free molecule and should display a notable intensity in case of non-parallel adsorption geometries. From the relative intensity ratios of these vibrational bands, the TiOPc adsorption geometry and orientation of the molecular 'plane' is thus readily derived. Specifically, and at variance to the study of Wei et al.,[11] our data unambiguously reveal a flat-lying configuration of the absorbed TiOPc molecule on the Ag(111) surface with an oxygen-up geometry. Accompanied STM and SPA-LEED data fully support this revised structure model of the TiOPc/Ag(111) monolayer.

Additional density functional theory (DFT) calculations have been carried out in order to search for the most stable absorption configurations of the TiOPc molecules on the Ag(111) surface. These DFT results, together with the SPA-LEED and STM findings, allow us to develop a structure model for the commensurate monolayer of TiOPc on Ag(111) where the molecules are subject to high packing densities; this is accomplished by an azimuthal rotation of adjacent molecules in order to minimize the repulsive intermolecular forces. Furthermore, although the charge transfer towards the molecule in the DFT calculations is rather small, the calculated position of the lowest unoccupied molecular orbital (LUMO) close to the Fermi level is consistent with the presence of interfacial dynamical charge transfer associated with the excitation of certain vibrational modes, explaining their non-negligible IR intensities.

## 2. METHODS

### 2.1 Experimental section

The measurements reported here were performed in two different ultrahigh vacuum (UHV) systems. Spot-profile analysis low energy electron diffraction (SPA-LEED) and Fourier transform infrared absorption spectroscopy (FT-IRAS, Bruker IFS 66v) with evacuated optics (p < 10 mbar) were performed in a system described in detail elsewhere.[13] IRAS measurements were performed at a sample temperature of 80K and 1000-2000 scans have typically been co-added at 2 cm$^{-1}$ spectral resolution.

The STM measurements were obtained in another chamber using a variable temperature scanning tunneling microscope (Omicron VT-STM) operated at 110 K in the constant current mode and using etched tungsten tips. For the present study different silver substrates have been used: a Ag(111) single crystal (Mateck, purity 5N), as well as Ag(111) films (150 nm) that were epitaxially grown under high vacuum conditions onto freshly cleaved and carefully degassed mica substrates. Prior to each organic film deposition, the silver surfaces were cleaned in situ by repeated cycles of Ar$^+$ sputtering ($E_{ion}$ = 800 eV) and annealing (700 K) until a sharp (1x1)-LEED pattern with a low background signal was observed and no traces of contaminations were found by STM or IRAS. The temperature of the substrates was measured using a K-type thermocouple attached directly to the sample, hence providing a precise temperature reading.

TiOPc thin films were prepared under UHV conditions by sublimation from a Knudsen cell. The evaporant was purified by extensive degassing at 550 K and analyzed by quadrupole mass spectrometry (Pfeiffer QMG 700, mass range 0 – 1024 u); a detailed analysis of the gas flow composition, before and after the degassing procedure, is depicted in Figures S6 and S7 of the Supporting Information. To enable formation of highly ordered films, deposition was performed at elevated substrate temperatures of 300 - 450 K and using low deposition rates of 0.05 monolayers/min. The deposition rate was estimated from IRAS measurements which reveal a clear indication of monolayer completion (saturation of monolayer bands and appearance of bilayer vibrational modes). For the STM measurements TiOPc films were grown at typical deposition rates of 0.5 Å/min, as determined by a quartz crystal microbalance. All matrix notations of LEED pattern have been verified by modelling using LEEDpat.[14]

## 2.2 Theoretical methods

Our DFT calculations were performed using the Vienna *ab initio* simulation package (VASP)[15] and the Perdew, Burke, Ernzernhorf (PBE) generalized gradient corrected exchange and correlation functional[16] with D3 Grimme corrections[17,18] to incorporate the effect of long-range van der Waals interactions. We used a 400 eV plane-wave cutoff as well as the projected-augmented-wave (PAW) method[19] to describe the atomic cores.

Our calculated lattice parameter for bulk Ag is 4.073 Å, in good agreement with the experimental value 4.056 Å.[20] A 25x25x25 Monkhorst-Pack k-point grid[21] was used in this bulk calculation to sample the Brillouin zone. Using this calculated bulk lattice parameter we constructed a slab containing 3 Ag layers, relaxing the topmost surface layer. Relaxations were pursued until forces acting on all the atoms were smaller than 0.03 eV/Å. We kept a large vacuum distance (20 Å) to avoid interaction between periodic replicas of the slab along the z-direction.

We then used a $\begin{pmatrix} 7 & 4\sqrt{3} \end{pmatrix}$ lateral supercell to study the interaction of the TiOPc molecules with Ag(111) in the low coverage regime, i.e., these calculation aim to explore mostly the effects of the molecule-substrate interaction. For these calculations we used different adsorption configurations as described in detail below and a 3x3x1 k-sampling.

## 3. RESULTS AND DISCUSSION
### 3.1 TiOPc/Ag(111) monolayer phases

In the present study, the TiOPc coverage ($\theta_{TiOPc}$) was systematically varied between 0.5 -1 monolayers (ML) and various distinct phases were identified depending on the amount of deposited TiOPc. The corresponding LEED patterns and STM micrographs are summarized in Figure 1, and reveal the evolution of the different phases. To reduce the Debye-Waller attenuation of the diffraction peaks and admolecular diffusion all LEED and STM measurements were performed at cryogenic temperatures (80 - 110 K). At low coverage ($\theta_{TiOPc}$ < 0.6 ML) a 2D-gas phase is found which, for increasing coverages, transforms into a long range ordered, commensurate adlayer structure (c-phase) and finally, upon completion of the monolayer, into a point-on-line (POL) phase. Similar sequences of long range ordered structures have been reported during the growth of other phthalocyanines like CuPc, H$_2$Pc and SnPc on coinage metal substrates.[13,22-25] Thus it seems that such monolayer structures are characteristic for adsorbed phthalocyanines. Beyond these similarities, TiOPc reveals, however, some particular features that we present below and which may be related to the specific interaction between molecule and substrate. The initially formed 2D-gas phase reveals a characteristic disc-like LEED pattern (cf. Figure 1a) that reflects random molecular locations, with the disc radius defining the minimum distance between the TiOPc molecules. A detailed analysis of the SPA-LEED measurements obtained at a coverage of 0.5 ML yields a minimum center-to-center distance of around 16 Å. Accompanying STM measurements (see Figure S1, Supporting Information) at this coverage reveal extended regions of a dilute molecular adlayer. Despite cooling to temperatures of 110 K the high molecular mobility hampered imaging with molecular resolution. When increasing the layer coverage above 0.6 ML a highly ordered phase that exhibits sharp diffraction spots (cf. Figure 1b) begins to form at low T, hence reflecting the distinct long-range order of the film. The diffraction pattern is identified as a $\left(7 \times 4\sqrt{3}\right)$ structure appearing in several rotational and mirror domains with respect to the high symmetry directions of the silver surface. Notably the rectangular unit cell of this TiOPc overlayer is commensurate with respect to the hexagonal Ag(111) surface and therefore is referred to as c-phase. In matrix notation this overlayer can be described as $\begin{pmatrix} 7 & 0 \\ 4 & 8 \end{pmatrix}$.

The unit cell area of this TiOPc c-phase amounts to 402 Å$^2$, which is very large compared to the unit cells reported for other Pc molecules, e.g. CuPc, or H$_2$Pc on Ag(111) that contain 1 molecule per unit cell and exhibit an area of approximately 217 Å$^2$.[23,25] Indeed, the accompanying constant current STM measurements show that the unit cell of the c-phase actually contains two inequivalent TiOPc molecules as depicted in Figure 1e so that the area allotted to each molecule amounts to 201 Å$^2$. Additional large area STM micrographs reveal the presence of straight step edges of the substrate. It is presumed that they correspond to the dense packed substrate rows of Ag(111) oriented along the $\langle 1\bar{1}0 \rangle$ azimuthal directions which allows us to derive further information regarding the azimuthal molecular orientation. Closer inspection of such data supports the idea that the two molecules within the unit cell display different azimuthal orientations and exhibit angles of their molecular diagonal with respect to the $\langle 1\bar{1}0 \rangle$ azimuth of 4° and 27°, respectively (see Figure S2, Supporting Information). We attribute the slightly different azimuthal rotations of the two inequivalent molecules to a competition between adsorbate - substrate and intermolecular interactions: (i) preference for particular sites and azimuthal orientations of the molecules, and (ii) repulsive interaction among adjacent molecules due to the high packing density of TiOPc.

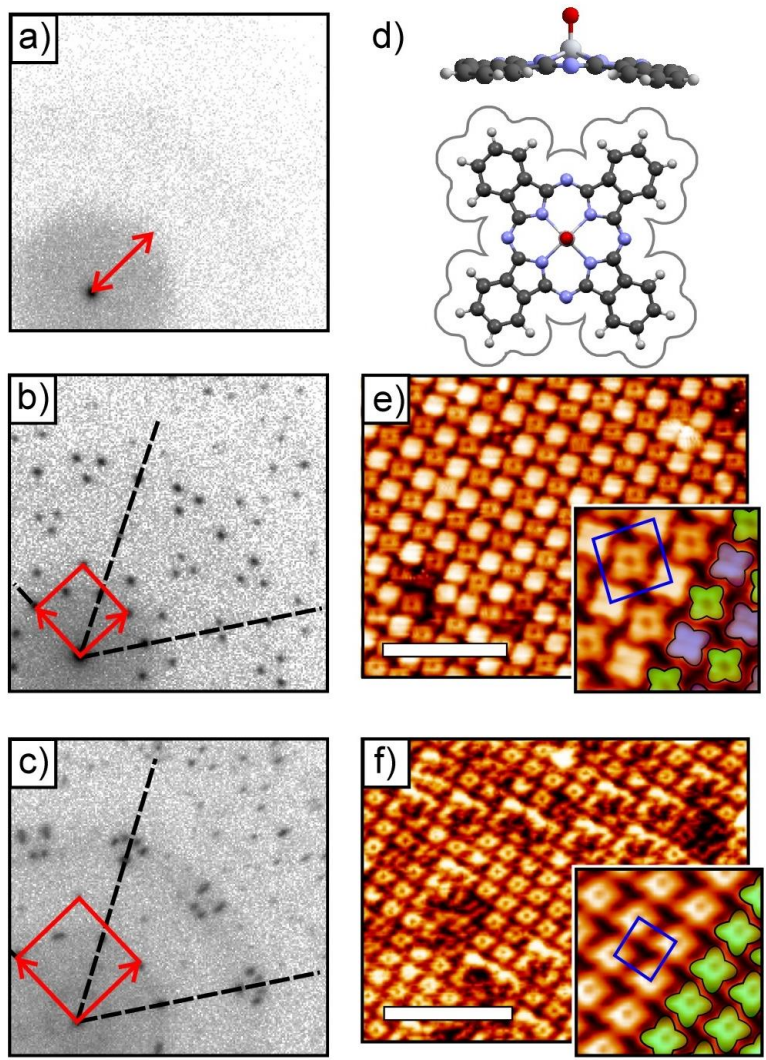

Figure 1: SPA-LEED images and corresponding STM micrographs of TiOPc monolayer phases grown on Ag(111). (a) SPA-LEED measurement of a 2D gas-phase with a disc-like pattern. The radius of the inner disc represents the minimum intermolecular distance and is marked in red. (b,e) commensurate phase (c-phase). In the SPA-LEED measurement the close-to-square surface unit cell and Ag(111) high symmetry directions are highlighted in red and black, respectively. The STM micrograph reveals the presence of two molecules within the unit cell. Both molecules display slightly different azimuthal orientations relative to each other. The unit cell orientation and molecular schemes are superimposed to the image ($U_{Bias}$ = 0.39 V, $I_t$ = 150 pA). (c,f) point-on-line phase. In the SPA-LEED measurement the surface unit cell and Ag(111) high symmetry directions are highlighted in red and black, respectively. The surface unit cell (real space) results to be about x0.5 smaller than in the c-phase. The STM micrograph reveals that only one molecule is contained in the surface unit cell ($U_{Bias}$ = 0.43 V, $I_t$ = 10 pA). (d) TiOPc molecule with corresponding van-der-Waals dimensions (top and side view). All SPA-LEED measurement were performed at 80 K ($E_{kin}$ = 50 eV). All STM micrographs were taken at T = 110 K; the inset micrographs represent correlated averages. The scale bars in (e) and (f) amount to 10 nm.

Further increasing the TiOPc coverage until completion of the monolayer leads to an abrupt change of the LEED pattern. From the diffraction pattern that is shown in Figure 1c the adlayer can be defined as a $\begin{pmatrix} 4.875 & -0.125 \\ 2.625 & 5.625 \end{pmatrix}$ superstructure. The corresponding unit cell implies a distinct arrangement of molecules which is referred to as a point-on-line (POL) phase. In such a POL-phase the fixed registry between molecules and the Ag(111) surface is lost, while the unit cell vectors seem to shift along the direction of substrate atom rows, e.g. the $\langle 1\bar{1}0 \rangle$ azimuth of Ag(111);[26] the molecules then arrange adequately in order to optimize their positions within the layer, i.e. by minimizing repulsive interactions among the closely spaced molecules. We note that at full saturation the relative importance of site preference and intermolecular repulsion has changed with respect to the c-phase and the latter becomes dominant.

Such phases are frequently observed in organic layers on weakly interacting metallic surfaces.[9,26] The POL-phase unit cell size (199 Å$^2$) is very similar to the (single) molecular area adopted in the c-phase at slightly lower coverages which suggests a primitive unit cell. In fact, the corresponding STM measurements clearly confirm such an arrangement and reveal a uniform azimuthal orientation with only a single molecule per unit cell (Figure 1f and Figure S3 of the Supporting Information). We note that the TiOPc foot print area in the (001) plane of the bulk phase reveals a molecular area of 193 Å$^2$,[27] which indicates that the molecules in the POL-phase are not most densely packed.

Even though the non-integer numbers within the above-mentioned matrix can be interpreted as the absence of a fixed registry between molecules and the Ag(111) surface, the found structure and matrix may actually be denoted as a high-order commensurate (HOC) phase.[28]

### 3.2 TiOPc vibrational properties and molecular orientation

An interesting feature associated with the STM measurements of the c-phase is the non-uniform imaging contrast for adjacent molecules (see Figure 1e). The same effect has been reported in the literature for a related system, namely vanadyl-phthalocyanine (VOPc) molecules grown on Ag(111);[29] there, the authors have explained this observation by the presence of two different molecular orientations of adsorbed VOPc molecules, i.e. O-down and O-up, that may coexist in the layer in direct contact with the Ag(111) substrate.

In order to clarify this issue, that is, which type of dipole alignment prevails within the TiOPc/Ag(111) layers at coverages up to completion of the full monolayer we have performed IRAS measurements. In Figure 2 vibrational spectra representative for the 2D-gas, commensurate and POL-phase of TiOPc/Ag(111) are displayed. The spectral range 700 - 1000 cm$^{-1}$ primarily encompasses modes with out-of-plane character with respect to the TiOPc aromatic backbone. Among them, two C-H vibrational bending modes located at the periphery of the Pc species are observed at about 715 and 764 cm$^{-1}$. The most characteristic feature of the displayed TiOPc spectra, however, is the prominent band at 993 cm$^{-1}$ that can be ascribed to the axial titanyl group (Ti=O stretching mode).[30] A closer inspection of this band reveals a peculiar line shape which can be well explained by isotopic splittings due to the natural abundance of Ti isotopes.[31] Despite this line splitting the Ti-O band is sharp, very well defined and exhibits a constant vibrational frequency in the entire coverage regime (0.5-1.0 ML) examined here, i.e. for all three different phases of adsorbed TiOPc. We take this as a clear indication for a unique Ti=O group orientation. Moreover, its intensity varies uniformly and gradually as the TiOPc coverage is steadily increased.

It is evident from Figure 2 that vibrational bands associated with adsorbed TiOPc change only slightly as the coverage increases from the 2D-gas to the POL-phase, except for an intensity increase and very minor frequency shifts due to changes in the local environment of the TiOPc molecules

Above 1000 cm$^{-1}$ some 'in-plane' vibrational modes of TiOPc are detected. The term 'in-plane' thereby assumes that the phthalocyanine frame represents a virtually flat entity, even though the TiOPc species shows a slight pyramidal shape. Specifically, weak in-plane vibrational modes are observed at about 1130, 1370 and 1491 cm$^{-1}$, the latter two displaying asymmetric line shapes. For molecules grown on metallic surfaces, like in the present case, only those vibrations with dynamic dipole moments perpendicular to the metal surface are detected due to the surface selection rule.[32] Due to the slight pyramidal shape of the Pc molecular frame for TiOPc, those in-plane modes which belong to the $A_1$ irreducible representation of the $C_{4v}$ symmetry group should actually be visible in IR spectra of TiOPc on Ag(111) even for a parallel adsorption geometry. Moreover, their weak intensity may be enhanced as a result of interfacial dynamical charge transfer induced by vibrational motion if a partially filled molecular orbital exists.[33-37] In accordance with similar modes found for CuPc/Ag(111),[13] they are ascribed to 'in-plane' deformation modes of the pyrrole and the benzene rings with an $A_1$ - type of symmetry. Thereby, the pronounced asymmetry is a clear indication for non-adiabatic processes governing these excitations and the associated coupling between vibrational and electronic motion.

For TiOPc on Ag(111) no further prominent in-plane modes are detected, apart from the mentioned weak in-plane modes; according to vibrational spectra of TiOPc/KBr-pellets (see Figure S8 in the Supporting Information) very prominent bands of the TiOPc species should, in particular, be present in the region 895, 1050-1075, 1280-1290, and 1330 cm$^{-1}$ for non-recumbently oriented molecules. Their absence unambiguously demonstrates that TiOPc has its molecular plane oriented parallel to the Ag(111) surface in the first monolayer.

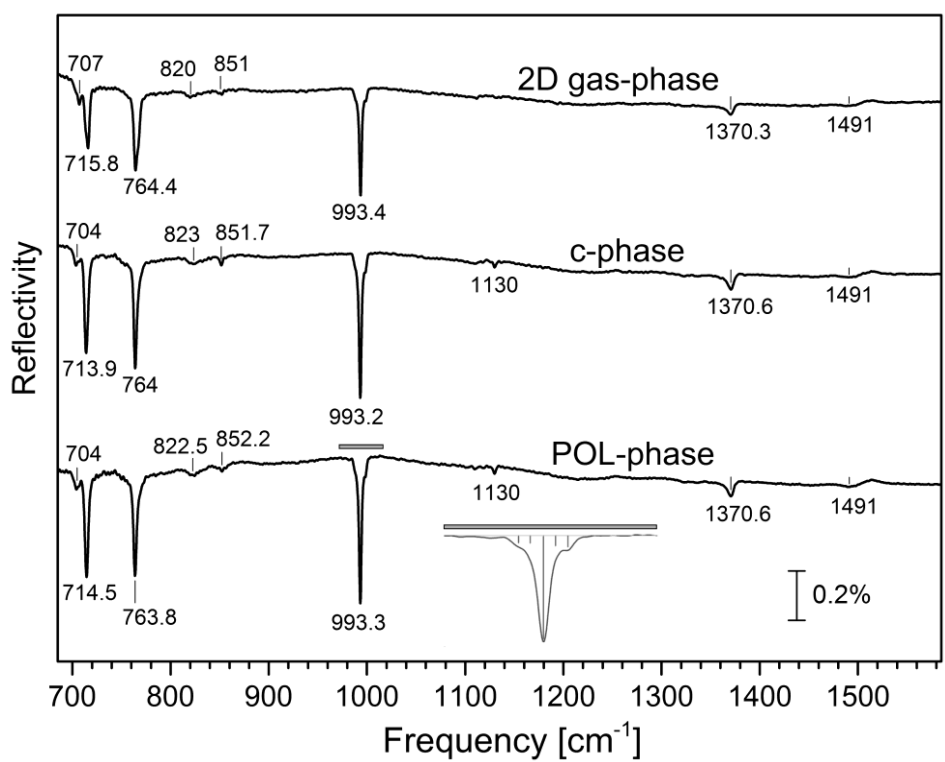

Figure 2: Coverage dependency of IR spectra of TiOPc adsorbed on Ag(111) in the coverage range 0.5 – 1 ML. The displayed spectral range covers out-of-plane vibrational modes (C-H bending modes and Ti=O stretching mode) at <1000cm$^{-1}$, as well as weak in-plane modes with asymmetric line shapes at higher frequencies. The Ti-O stretching band of the POL-phase at 993.3 cm$^{-1}$ has been enlarged to better display the isotopic splitting of this band, with the natural abundances of the various Ti isotopes indicated by the bars. All spectra have been obtained at 80 K, using a spectral resolution of 2 cm$^{-1}$.

On the basis of IRAS the precise orientation, i.e. O-down or O-up adsorption geometry of TiOPc can be derived only indirectly. Nonetheless these data provide unambiguous evidence which render the O-down configuration of TiOPc on Ag(111) implausible under the growth conditions in our study ($T_{growth}$ = 300-450 K, deposition rate 0.05 ML/min). In general, the van der Waals interaction between TiOPc and Ag(111) should strongly favour a geometry with short distances of the phthalocyanine molecular frame with respect to the Ag(111) surface, i.e. an O-up geometry. The O-down arrangement, on the other hand, is associated with a notably larger average vertical bonding distance of the π-conjugated backbone and should yield lower adsorption energies (considering van der Waals interaction alone). Only in case of a substantial chemical interaction of the titanyl oxygen with the silver substrate this choice/preference may get reversed (which requires a notable distortion of the Ag surface lattice). In fact, a chemical interaction of a hypothetical O-down TiOPc geometry should be accompanied by two evidences: (i) a chemical shift of the $O_{1s}$ photoemission peak in X-ray photoemission (which we cannot verify in our chamber) and (ii) a notable frequency shift of the Ti=O stretching mode with respect to the free molecule (which we can rule out based on our IR data).

Our DFT calculations seem to indicate that for some adsorption sites of TiOPc (e.g., bridge sites), O-down molecules compete in stability with the O-up TiOPc on Ag(111). However, as discussed below, the experimental evidence seems to rule out the presence of an O-down configuration under the present growth conditions. Thus, we are inclined to ascribe this result to the limitations of the density functional utilized here to describe the delicate competition between the long-range van de Waals interactions with the substrate and the short-range O-Ag chemical bond.

Representative frequencies for a free TiOPc molecule can be derived from bulk measurements (TiOPc/KBr-pellets or grown thick films)[38] or theoretical calculations of a free molecule,[30] where the effect of a substrate is not present. In the literature, the Ti=O stretch mode $v_{Ti-O}$ has been identified at frequencies of 971 cm$^{-1}$ (bulk sample), 963 cm$^{-1}$ (TiOPc-KBr pellet), and 1057 cm$^{-1}$ (calculation), that are relatively close to our experimentally observed Ti=O stretch mode frequency (993.3 cm$^{-1}$) for the TiOPc/Ag(111) monolayer. The Ti=O group obviously vibrates pointing towards the vacuum and we conclude that a TiOPc species which may be chemically bound to the Ag(111) surface via the oxygen is not present. Moreover, we can definitely discard the simultaneous presence of both O-down and O-up molecular orientation in the sub-monolayer and monolayer growth of TiOPc on Ag(111): a mixed O-up and O-down configuration for the c-phase which converts to a uniform O-up POL-phase at slighty higher $\Theta_{TiOPc}$ should lead to an abrupt increase of the Ti=O band as well as the disappearance of the mode associated with an O-down configuration, which is both not observed.

Finally, one still may invoke that a tilted oxygen-down species may be present; such a distorted adsorption geometry has been suggested based on STM data.[11,39] In this respect we note that a tilted molecular geometry has already been excluded based on the lack of prominent in-plane vibrational modes of TiOPc/Ag(111).

The minor difference between our measured $\nu_{Ti-O}$ frequency (adsorbed TiOPc) with respect to a free TiOPc molecule indicates further that TiOPc adsorbed on a Ag(111) surface is rather weakly chemisorbed, which means that its electronic structure and vibrational properties should be only moderately perturbed. The relatively small perturbation of the molecular electronic structure when the molecule adsorbs in the O-up parallel configuration is further confirmed by the DFT calculations (see below). The prevalence of van der Waals interactions is not surprising and it is in accordance with similar conclusions derived in studies of various Pc molecules on Ag(111) surfaces.[24,40]

### 3.3 Impurity-dominated Adlayers

We like to mention that in the course of this study various preparation conditions have been compared, revealing in particular another type of adlayer structure in the STM measurements. Using the commercial TiOPc raw material without elaborate purification and with degassing of the Knudsen cell for a few hours only, a notable flux of about 0.6 Å/min was obtained already at crucible temperatures of 540 K. Depositing the evaporating molecules onto a Ag(111) surface at room temperature yields a well ordered adlayer. As depicted in Figure 3b, the corresponding STM data exhibit characteristic triangularly shaped molecules that form a nearly hexagonal lattice. The unit cell vectors (13.2 Å) form an angle of about 60° hence yielding a molecular area of 151 Å$^2$.

After extensive degassing, the initial flux could no longer be maintained at this low cell temperature and instead required substantially higher cell temperatures of about 620K. Deposition of such carefully degassed TiOPc material, by contrast, yields a distinctly different monolayer film structure as depicted in Figure 3a. Here, a nearly rectangular unit cell with an area of about 200 Å$^2$ is observed which is equivalent to the monolayer films presented in Figure 1, and which have been prepared from likewise purified TiOPc.

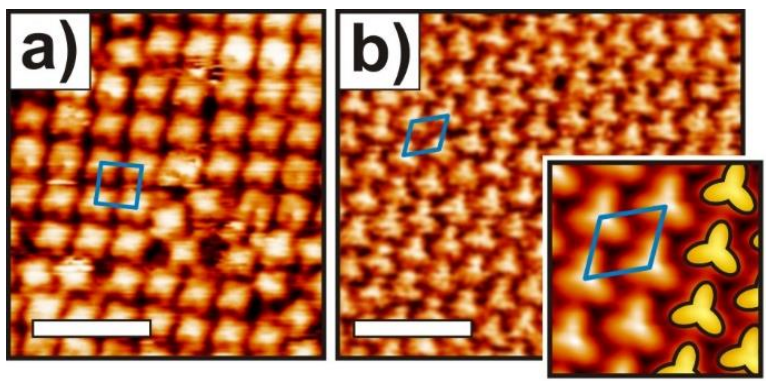

Figure. 3: STM micrographs that compare two monolayer films prepared by deposition of a) extensively purified TiOPc (U$_{bias}$=-0.95V, I$_t$ = 50pA) and b) only mildly degassed TiOPc raw material (U$_{bias}$=0.35V, I$_t$ = 380pA). Scale bar: 5nm.

An almost identical adlayer structure, with characteristic triangularly shaped molecules and essentially the same unit cell size (152 Å$^2$), was reported before for a nominal TiOPc monolayer on Ag(111).[11,39] According to the authors, a non-planar, inclined adsorption geometry of TiOPc molecules prevails, leading to the triangular appearance of the TiOPc species instead of a quadratic look expected for flat lying oriented molecules. Interestingly, a very similar triangular shaped adlayer structure was observed in a STM study for a monolayer of chloroboron-subphthalocyanine on Au(111).[41]

To demonstrate the difficulty to obtain clean TiOPc powder lacking volatile contaminants we have also used mass spectrometry to monitor and analyze the molecular flux at different stages of degassing (cf. Figures S6 and S7 in the Supporting Information). Initially, these data exhibited a notable signal at m/z=384u corresponding to the mass of the subphthalocyanine-fragment ($C_{24}N_6H_{12}$), while its intensity vanished upon continued degassing. Moreover, complementary IRAS data that were recorded for films of little degassed TiOPc grown onto KBr pellets exhibit distinctly different vibrational spectra than found for properly purified TiOPc and, in particular, showed no Ti-O stretching mode (cf. Figure S8, Supporting Information); this suggests that the film shown in Figure 3b does not represent an intact TiOPc.

In this respect it is interesting that Liu et al.[39] used a crucible temperature of 500 K to grow their layers that agrees with the temperatures used in our experiment leading the similar triangular shaped species (see Figure 3b). According to our observations related to the additional degassing of the TiOPc evaporant, the prevalent surface species in Figure 3b and in the Refs.[11,39] seems well defined, but comes with an identity other than TiOPc.

### 3.4 TiOPc /Ag(111) adsorption sites

In order to investigate the adsorption of TiOPc on a Ag(111) surface in more detail, we resort to density functional theory (DFT) calculations to search for preferential adsorption sites on the substrate. Various molecular adsorption configurations were studied with the central Ti atom located above fcc, top and bridge positions of the Ag surface. Additionally, the azimuthal alignment of the TiOPc molecule relative to the substrate rows was examined to derive the preferred molecular arrangement, and considering two different azimuthal orientations: (i) with two of the inner-ring N atoms aligned along the $\langle 1\bar{1}0 \rangle$ high symmetry direction of Ag(111), in the following referred to as top$_+$ and bridge$_+$, and (ii) a 45° rotated geometry of TiOPc with respect to the $\langle 1\bar{1}0 \rangle$ high symmetry direction of Ag(111), which will be termed as top$_x$, bridge$_x$, and fcc$_x$.

|  | $E_{ads}$ (eV) | $h_1$ (Å) | $h_2$ (Å) |
| --- | --- | --- | --- |
| gas phase |  | 0.707 ± 0.009 |  |
| fcc$_x$ | 3.663 | 0.643 ± 0.011 | 3.294 ± 0.023 |
| top$_x$ | 3.687 | 0.656 ± 0.009 | 3.292 ± 0.022 |
| top$_+$ | 3.538 | 0.643 ± 0.009 | 3.299 ± 0.025 |
| bridge$_x$ | 3.688 | 0.653 ± 0.009 | 3.292 ± 0.022 |
| bridge$_+$ | 3.741 | 0.645 ± 0.009 | 3.289 ± 0.020 |

Table 1: Calculated quantities for TiOPc/Ag(111) in the low coverage regime described by a single molecule adsorbed on a $(7 \times 4\sqrt{3})$ supercell along the surface and different adsorption geometries.

$E_{ads}$: adsorption energy per molecule, $h_1$: distance (including its standard deviation) between Ti atom and the TiOPc molecular plane (aromatic backbone); $h_2$: distance (including its standard deviation) between the Ag (111) surface and the TiOPc molecular plane.

The calculated adsorption energies ($E_{ads}$) and molecular heights for the different adsorption sites of TiOPc are summarized in Table 1. To mimic the situation at low coverages, a single molecule arranged in a $\left(7 \times 4\sqrt{3}\right)$ supercell of the Ag(111) surface has been computed. It is found that there is a weak preference for the bridge₊ adsorption geometry with an $E_{ads}$ of 3.741 eV, followed by bridge$_x$ (3.688 eV) and top$_x$ (3.687 eV) configurations. In accordance with the minor variations in adsorption energy, the differences in the adsorption geometry are rather small as well.

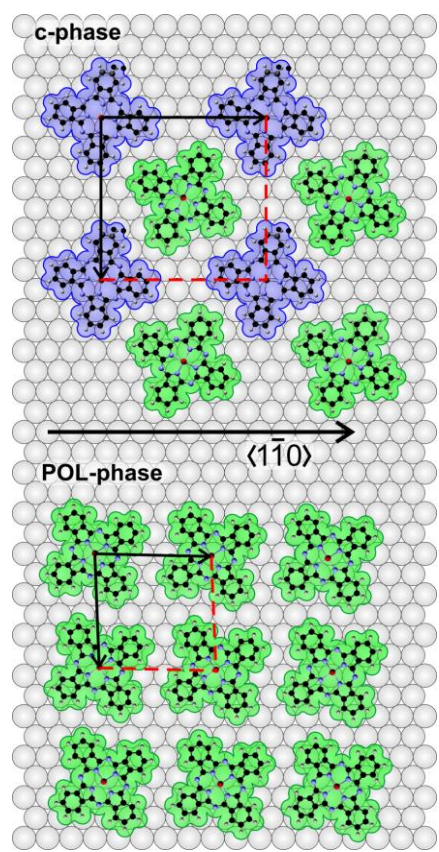

*Figure 4: Structure models of the molecular arrangement derived from SPA-LEED and STM data and DFT calculations for the c-phase and POL-phase. Unit cell dimensions are 201 Å² per molecule for c-phase and 199 Å² for POL-phase.*

Considering our DFT results about the most stable molecular adsorption sites on Ag(111) as well as the corresponding STM measurements of the c-phase (cf. Figure 1e), we suggest a model for the molecular arrangement of TiOPc molecules in the c-phase which is depicted in Figure 4. There we assume that the central Ti atoms of the molecular layer are located on well-defined adsorption sites, i.e. bridge or top positions of the Ag(111) surface, while the azimuthal orientation of the molecules is adjusted according to the observed STM measurements (cf. Figure 1e and Figure S2 of the Supporting Information). This can also explain the different molecular contrast observed in the STM data. We excluded the top$_+$ sites as the associated adsorption energy is notably lower as compared to the alternative sites. Even though the calculated adsorption geometries refer to well defined azimuthal orientations with respect to the Ag(111) $\langle 1\bar{1}0 \rangle$ azimuth, we allowed the molecules to rotate slightly in order to qualitatively account for the repulsive intermolecular interactions among adjacent molecules.

Upon completion of the monolayer all molecules reveal a rather uniform azimuthal orientation. Though the molecules occupy in this phase bridge-like adsorption sites which are energetically slightly less favoured this enables a higher packing density and allow for the formation of a high order commensurate POL-phase.

In accordance with the weak dependence of the adsorption energy on the adsorption site, the electronic structure of the various adsorption configurations is likewise found to be almost identical. Partial density of states (DOS) calculations performed for the system TiOPc/Ag(111) for the different adsorption sites described above do not display substantial differences as shown in Figure 5. This proves directly that the orbital distributions of the molecule are not affected by the specific adsorption site.

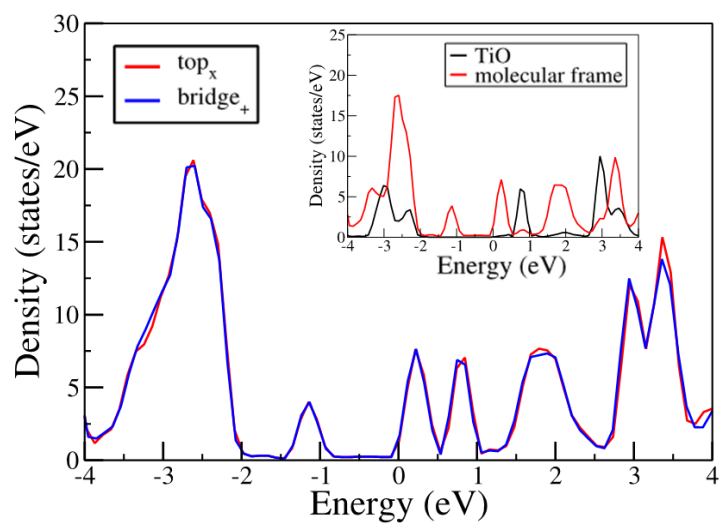

Figure 5: DOS projected onto the molecule calculated for two different adsorption configurations of TiOPc on Ag(111). The inset shows the DOS projected onto the titanyl group and the molecular backbone for the case of the bridge$_+$ sites. Energies are referred to the Fermi level. Notice the presence of the LUMO close to the Fermi level, although with an almost negligible population, in contrast to the case of other metalized Pc molecules that present a partially filled LUMO.

As mentioned above, O-down molecules were also examined and, for bridge sites, they were found to be marginally more stable than O-up molecules. However, due to the plethora of experimental evidences that rule out the presence of O-down molecules in the present experiments, we only focus here on the O-up configuration. An important point which may help to understand why the O-up configuration is the only one that is observed upon room temperature TiOPc film preparation concerns the strong deformations of the Ag(111) lattice associated with the O-down configuration. Specifically, the calculated high adsorption energy of such species involves displacements of surface Ag atoms from their original sites. As incoming TiOPc interacts with a pristine Ag(111) lattice it cannot take advantage of the calculated energy gain. Rather, there will be a barrier associated with deformation of the Ag(111) substrate and which probably prevents TiOPc to reach an O-down geometry, even if energetically possible. The time scale required to produce such a deformation thus competes with the short interaction time of incoming TiOPc to thermalize.

In this context it should be mentioned that several STM studies have reported the occurrence of different molecular orientations in thin films of axially functionalized phthalocyanines.[42-47] In particular they observed an STM tip mediated orientational switching of the axial group by applying short voltage pulses. In the present study this process can be excluded because rather gentle methods such as IRAS, SPA-LEED and low current STM at low bias voltage were used.

In addition, the adsorption-induced charge rearrangements for the different hypothetical adsorption sites have been determined, yielding very similar results with minor variations. In all cases we observe a redistribution of the electron density in the proximity of the molecular backbone with a slight tendency to accumulate charge density right above the top metal layer, as it is depicted in Figure 6. Interestingly, electron density accumulation in the vicinity of the top metal layers is observed primarily in the surface regions not covered by the organic molecule. We attribute this response of the charge density to the Pauli pushback, commonly observed at all metal-organic interfaces.[48-50]

To obtain a more quantitative picture we also calculate the plane-integrated charge density rearrangement, $\Delta\rho(z)$ (Figure 6c), which apparently shows a weak dependency on the type of adsorption site, showing for all the considered geometries a characteristic electronic density oscillation in the molecule due to the high dipole moment associated with the TiO group. Using the calculated induced electronic density, $\Delta\rho(z)$, the dipole moment along the z direction was evaluated and the results summarized in Table 2; for the various calculated adsorption geometries only small differences in the total induced dipole moments are found. Therefore only slight differences regarding the work function changes are expected for the various adsorption sites and configurations.

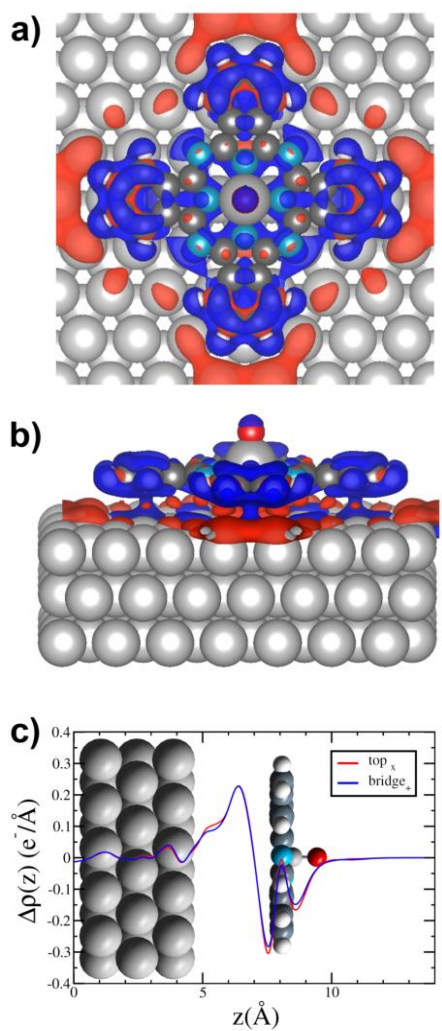

Figure 6: Induced electronic density for TiOPc adsorbed at a bridge$_+$ site with the molecule aligned along the $\langle 1\bar{1}0 \rangle$ symmetry direction of the Ag(111) surface. (a) Top view of the isosurface (isovalue = ± 2x10$^{-4}$ e/Bohr$^{-3}$) of the induced electronic charge; red and blue surfaces indicate respectively electron accumulation and depletion. (b) Same as (a) but side view. (c) Induced electronic density $\Delta\rho(z)$ averaged in the xy plane and plotted along z. The induced electronic charge is defined as the difference between the electronic charge of the total system and the electronic charge of the subsystems (molecule and metal substrate) with the geometries obtained after relaxation.

| TiOPc configuration | $d_{mol}$ (eÅ) | $d_{ind}$ (eÅ) | $d_{total}$ (eÅ) |
|---|---|---|---|
| gas phase | -0.59 | -- | -- |
| $fcc_x$ | -0.62 | 0.94 | 0.28 |
| $top_x$ | -0.61 | 0.94 | 0.29 |
| $top_+$ | -0.62 | 0.89 | 0.23 |
| $bridge_x$ | -0.62 | 0.96 | 0.30 |
| $bridge_+$ | -0.63 | 0.91 | 0.24 |

Table 2: Calculated vertical dipole moments of adsorbed TiOPc. The molecular dipole $d_{mol}$ in the second column corresponds to the dipole of the molecule computed using the molecular geometry obtained after adsorption. The third column shows the induced dipole $d_{ind}$ computed using the charge density distributions shown in Figure 6c. Finally, the fourth column shows the total dipole $d_{total}$ per unit cell of our slab containing one adsorbed molecule on one side.

Moreover, we have applied the Helmholtz equation to calculate the work function change induced by a deposited TiOPc monolayer on Ag(111). According to Table 2, the average total dipole per adsorbed molecule along the z-direction amounts to ≈ 0.28 eÅ: By using an area of 200 Å$^2$ per molecule, this yields a decrease of the work-function of the Ag(111) surface by 0.25 eV. This value agrees favorably with experimental observations using two-photon-photoemission spectroscopy.[51]

## 4. SUMMARY AND CONCLUSIONS

Using IRAS, STM and LEED measurements, together with DFT calculations, the adsorption of TiOPc on Ag(111) has been investigated in the present study. From IRAS we have unambiguous evidence that TiOPc is adsorbed in an O-up configuration with its molecular 'plane' oriented parallel to the surface at all coverages up to completion of the monolayer. This finding, while contradicting previous studies of TiOPc on Ag(111), is corroborated by our STM and LEED results. DFT calculations also found O-up molecules as the most stable adsorption geometry for the $top_x$ configuration. On bridge sites we found that O-up and O-down molecules are nearly degenerate (within ~30 meV per molecule). However, in view of the unambiguous experimental evidence presented here that rules out the presence of O-down molecules under the present growth conditions, we think that this result shows the limitations of the PBE-D3 functional to fully account for the subtle balance between van der Waals and chemical interactions in the present system. Moreover, energy barriers associated with distortion of the Ag(111) surface for O-down configurations might render such adsorption geometries inaccessible to impinging TiOPc.

We have experimentally identified three different phases that form during the growth of the first TiOPc monolayer. Besides a disordered 2D gas phase at low coverages, long range ordered phases (a commensurate and a POL-phase) are found. The latter is formed only at relatively high molecular packing density, i.e. when approaching saturation of the monolayer. In general, the molecules display an azimuthal misalignment with respect to adsorbate unit cell that helps to reduce the lateral repulsion between adjacent molecules at short distances. For the commensurate phase two inequivalent molecular adsorption sites have been identified which display different imaging contrast in STM as well as dissimilar azimuthal orientations. For the POL-phase all TiOPc molecules look identical in STM and they display a uniform azimuthal orientation.

According to the results of DFT, the molecules adsorbed with their center (i.e. Ti atom) above bridge sites and aligned along the $\langle 1\bar{1}0 \rangle$ direction of Ag(111) (bridge$_+$) represents the most favored adsorption geometry, while molecules with their centers above on-top or bridge sites and displaying a 45° azimuthal rotation with respect to the $\langle 1\bar{1}0 \rangle$ direction are found to be slightly less stable. On the basis LEED, STM and the adsorption energies derived by DFT, we suggest a structure model for the c-phase of TiOPc on Ag(111).

Comparison of TiOPc vibrational spectra with CuPc/Ag(111)[13] yields much weaker in-plane vibrational modes. As these modes acquire intensity primarily due to interfacial dynamical charge transfer between the Ag substrate and molecular electronic levels at the fermi level ($\varepsilon_F$) we conclude that the density of molecular states at $\varepsilon_F$ is substantially lower for TiOPc as compared to CuPc, which is in accordance with our DFT calculations and existing literature.[40]

Interestingly, the dipole moment associated with TiOPc adsorbed on Ag(111) is dramatically reduced as compared to the free TiOPc molecule. Moreover, the sign has changed which is attributed to a strong contribution induced by adsorption (Pauli push-back effect), and which is directed oppositely to the dipole associated with the axial Ti=O group of upright standing TiOPc. The resulting small vertical dipole is probably one of the causes that the c-phase is adopted over a wide range of coverages. By contrast, the larger dipole-dipole interaction (repulsion) for CuPc/Ag(111) leads to quasi-continuous coverage dependent intermolecular distances.[23] Since the repulsive interaction of parallel oriented dipoles is long ranging while the (lateral) Pauli-repulsion is active only at short distances the ordering of the various Pc species may be similar close to saturation of the monolayer, while exhibiting quite different behavior for dilute layers.

Based on QMS and IRAS data obtained after different stages of purification of the TiOPc raw material we like to emphasize the importance of careful degassing in order to avoid deposition of volatile impurities and/or molecular fragments.

ASSOCIATED CONTENT

**Supporting Information:**
Additional STM, QMS and IRAS data of TiOPc films on Ag(111).
**This material is available free of charge via the Internet at http://pubs.acs.org.**

**Author Information**
Corresponding Author
* peter.jakob@physik.uni-marburg.de : Tel +49 6421 28-24328
Notes: The authors declare no competing financial interest.

**Acknowledgments**
We gratefully acknowledge support from the Deutsche Forschungsgemeinschaft DFG (Germany) through the collaborative research center "Structure and Dynamics of Internal Interfaces" (SFB 1083, Projects A2, A3 and GP1).

involves negligible displacements within the phthalocyanine backbone. For such a simple system the frequencies $\nu_{\delta m}$ of the satellite bands associated with individual Ti isotopic species ($\delta m = \pm 1u, \pm 2u$) are readily calculated according to

$\nu_{\delta m} = \nu_0 \sqrt{\frac{m_{eff,0}}{m_{eff,\delta m}}}$ with $\frac{1}{m_{eff}} = \frac{1}{m_{Ti}} + \frac{1}{m_O}$ and $m_{eff,0}$ referring to the effective mass associated with the most abundant $^{48}$Ti isotope ($\delta m = 0$). For $\nu_0 = 993$ cm$^{-1}$ the shift amounts to about $\pm 2.5$ cm$^{-1}$ ($\delta m = \pm 1u$). A line shape analysis is provided in Figure S5 of the Supporting Information.

---

**TOC graphic:**

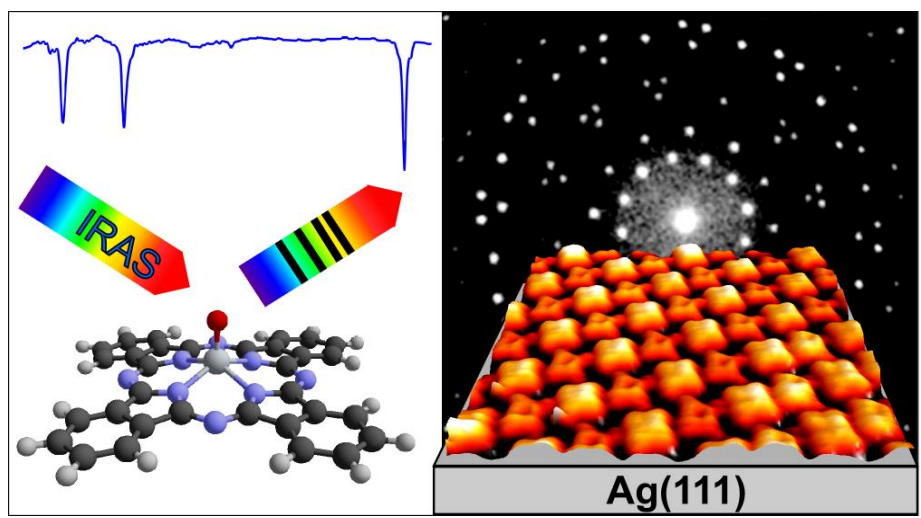